\documentclass[12pt,preprint,letterpaper]{emulateapj}
\begin{document}
\title{Measuring Transverse Motions for Nearby Galaxy Clusters.}
\author{Erika T. Hamden, Christine M. Simpson, Kathryn V. Johnston
\& Duane M. Lee}
\affil{Astronomy Department, Columbia University, New York, NY 10027, USA}

\begin{abstract}
Measuring the full  three-dimensional motions of extra-galactic objects in the Universe presents a seemingly insurmountable challenge. 
In this paper we investigate the application of a technique to measure tangential motion that has previously only been applied nearby within the Local Group of galaxies, to clusters of galaxies far beyond its borders.
We show that mapping the mean line-of-sight motion throughout a galaxy cluster could in principle be used to detect the {\it perspective rotation} induced by the projection of the cluster's tangential motion into the line-of-sight.
The signal will be most prominent for clusters of the largest angular extent, most symmetric intrinsic velocity distribution and surveyed with the largest number of pointings possible.
We investigate the feasibility of detecting this signal using three different approaches: 
measuring line-of-sight motions of individual cluster members; 
taking spectra of intracluster gas; and mapping distortions of the Cosmic Microwave Background radiation.
We conclude that future spectroscopic surveys of 1000's of members of nearby galaxy clusters hold the most promise of measuring cluster tangential motions using this technique.
\end{abstract}

\keywords{galaxies: clusters: general}
\section{Introduction}

It is our unhappy luck that even though we live in three dimensions, we are forced to view most of the Universe as a two dimensional projection with only one dimension of velocity information.  As such, we know very little about the motions of the things around us in any direction other than radially outward.  A more comprehensive view of motions in the Universe would provide a unique probe into the history and future of the things around us.  
Indeed, within the Milky Way, the HIPPARCOS \citep{perryman97} and near-future GAIA \citep{perryman02}  missions have brought about a rebirth in the ancient field of astrometry, pushing accuracies for direct proper motion measurements down first to mas/year precision and then on to tens of $\mu$as/year. 
The prospect of billions of stars with full 6-D phase-spa<ce motion known for an appreciable fraction of the Galaxy has allowed astronomers to propose measuring Galactic structure \citep[e.g.][]{johnston99} and recovering Galactic history \citep[e.g.][]{helmi00} in unprecedented detail.

Similar measurements of the full  phase-space positions of objects throughout and beyond the Local Group would allow an analogous reconstruction of the masses and past interaction of these objects and test the existence of large-scale flows in the Universe \citep[see][for an investigation on Local Group scales]{shaya03}.
Cosmological simulations of structure formation suggest that the peculiar velocity distribution of galaxy clusters has root-mean-square spread of $\sigma_{\rm pec} \sim$ 500 km/s \citep{sheth01}, corresponding to proper motion scales for galaxy clusters at distance $d$ from us of 
\begin{equation}
\mu \sim \left({v_{\rm tan} \over 500 {\rm km/s}}\right) \left({10 {\rm Mpc} \over d}\right) 10 {\mu {\rm as/year}}.
\label{eq:pm}
\end{equation}
At first sight, this suggests that the next generation of astrometric missions would have sufficient precision to detect these transverse motions.
However, prospects for direct measurements remain dim both because individual galaxies are extended and galaxy clusters have large internal velocity dispersions.

As an alternative to direct measurements of proper motions of nearby {\it stellar} clusters, Galactic astronomers have traditionally employed a neat geometrical trick that allows them to infer the motion of an object in three-dimensions from line-of-sight velocities alone. The method relies on the fact that the projection of the transverse motion into the line-of-sight will induce a gradient in the average line-of-sight velocities measured across any extended object
 --- an effect known as known ``perspective rotation'' \citep[hereafter PR ---][]{1961Feast}. 
PR has been used to verify measurements of the proper motion of the globular cluster Omega Centauri \citep{1997Merritt}, as well as the distance to the Large Magellanic Cloud \citep{2000Gould}.  
Most recently, \citet{2008Kaplinghat} pointed out that
there are now sufficient spectra taken for stars in nearby dwarf spheroidal galaxies for a precision of $\sim$ 100 km/s in tangential velocity estimates to be possible. \citet{walker08} subsequently verified this assertion with estimates for the motion of Fornax and Carina that agreed with prior astrometric measurments and the first three-dimensional measurement of the motion of Sextans. They also made estimates for Sculptor's proper motion that disagreed with prior work --- this disagreement could be explained as contamination by Sculptor's intrinsic rotation. 

To date, M31 is the most distant object that PR has been measured for \citep[using the line-of-sight velocities of 17 of its satellite galaxies, see][]{2008VanDerMarel} --- and this study represents over an order-of-magnitude leap in the distance to which such a measurement had been attempted. 
However, there is no limit in principle to the distance of objects for which this technique could prove useful. Indeed, many nearby clusters of galaxies have angular sizes of the same order-of-magnitude as M31's satellite system, with measurements of line-of-sight velocities of hundreds of members already taken and of ever increasing numbers of objects feasible in the near future.

In this paper, we examine to what extent we can expect to map bulk motions in the nearby Universe using current or near-future capabilities.
We review the technique of PR in more detail in \S 2, discuss 3 different types of data that should contain PR signatures in \S 3 and summarize our conclusions in \S 4.

\section{Approach: perspective rotation and galaxy clusters}

Consider an object, moving with radial and tangential velocities $(v_{\rm sys, rad}, v_{\rm sys,tan})$ relative to an observer.
The method of PR relies on the fact that the line of sight velocity $v_{\rm los}$ measured at any position 
in this object will contain some contribution from the
projection of both these systematic motions.  
Looking directly at the center of an object, the line of sight is perpendicular to $v_{\rm sys, tan}$, so $v_{\rm los}=v_{\rm sys, rad}$.  However, at any other position the angle the line of sight makes with the transverse motion is no longer $90^\circ$, 
and some portion of $v_{\rm sys, tan}$ will be projected onto the line of sight.  As the the angular separation $\theta$ of the line-of-sight from the center increases, this effect becomes more pronounced --- the total line-of-sight velocity at  angular separation $\theta$ and azimuthal angle $(\phi-\phi_{\rm tan})$  to the direction of tangential motion can be expressed as  
\begin{eqnarray}
\label{eq:LOSvel}
v_{\rm los}(\theta, \phi)=v_{\rm sys, los} \cos\theta &+& v_{\rm sys, tan} \sin\theta \cos (\phi - \phi_0) 
\cr
&+& v_{\rm local}(\theta, \phi),
\end{eqnarray}
where $v_{\rm local}$ represents the local mean motion within the object.

In the case of a solid, non-rotating object (i.e. where $v_{\rm local}(\theta, \phi)\equiv 0$), the three unknowns ($v_{\rm sys, rad}, v_{\rm sys,tan}$ and $\phi_{\rm 0}$) in equation (\ref{eq:LOSvel})  can be solved for given any three measurements of $v_{\rm los}$.
More realistically, we expect $v_{\rm local}$ to be non-zero due to a combination of random motions and rotation internal to the system, both of which could in principle be solved for in addition given an object of sufficient angular size measured at many positions  \citep[i.e. to maximize the signal, as done by ][for the LMC]{vandermarel02}. Note in particular, that a system does not need to be relaxed and in equilibrium with a maxwellian velocity distribution for PR to be apparent, though ideally it would have a fairly symmetric intrinsic velocity distribution and contain only limited substructures.

As noted in the introduction, PR has typically been used in the past to find transverse motions of Local Group objects that are degrees across and might be expected to have relative motions of 
100-300 km/s. 
In contrast, clusters of galaxies should have transverse velocites of order the expected root-mean-square peculiar velocity or $v_{\rm sys, tan}\sim \sigma_{\rm pec}$
\citep[where $\sigma_{\rm pec} \sim 500$km/s according to linear theory --- see][]{sheth01}.
The nearest galaxy clusters have angular radii of up to $\theta_{\rm max}=5^\circ$  (for the Virgo Cluster).
Hence, from the second term in equation (\ref{eq:LOSvel}) we can expect to be looking for a signal of a velocity difference  across the galaxy cluster of order 
\begin{equation}
\label{eq:signal}
\Delta v \sim {v_{\rm pec} \over 500 {\rm km/s}} \times { \theta_{\rm max}  \over 5^\circ } \; 180 {\rm km/s}.
\end{equation}

This ``signal'' can be compared to the expected sources of confusion. First, galaxies in galaxy clusters have a range in orbital velocities of order $\sigma \sim $ 500-1000 km/s (depending on the galaxy cluster's mass). 
They could also be rotating as a system due to tidal torques on the galaxy cluster as a whole --- \citet{cooray02} estimate rotation amplitudes of order 36-180 km/s at the galaxy cluster virial radius using linear theory \citep{peebles69}.
Lastly, they could have much larger apparent rotation due to substructure induced from recent off-axis mergers, or accretions of smaller groups \citep{ricker01}.

These considerations suggest that the most interesting galaxy clusters in which to look for PR would be those: (i) for which large numbers of velocity measurements could be made to beat down the uncertainty due to $\sigma$; (ii) that are nearby enough for their virial radii to subtend a large angle (of order a degree or more) so that the PR signal is not much smaller than that due to intrinsic rotation; and (iii) that are not clearly substructured.


\section{Results: measuring perspective rotation using....}

\subsection{.... cluster galaxies.}

\begin{figure}
\plottwo{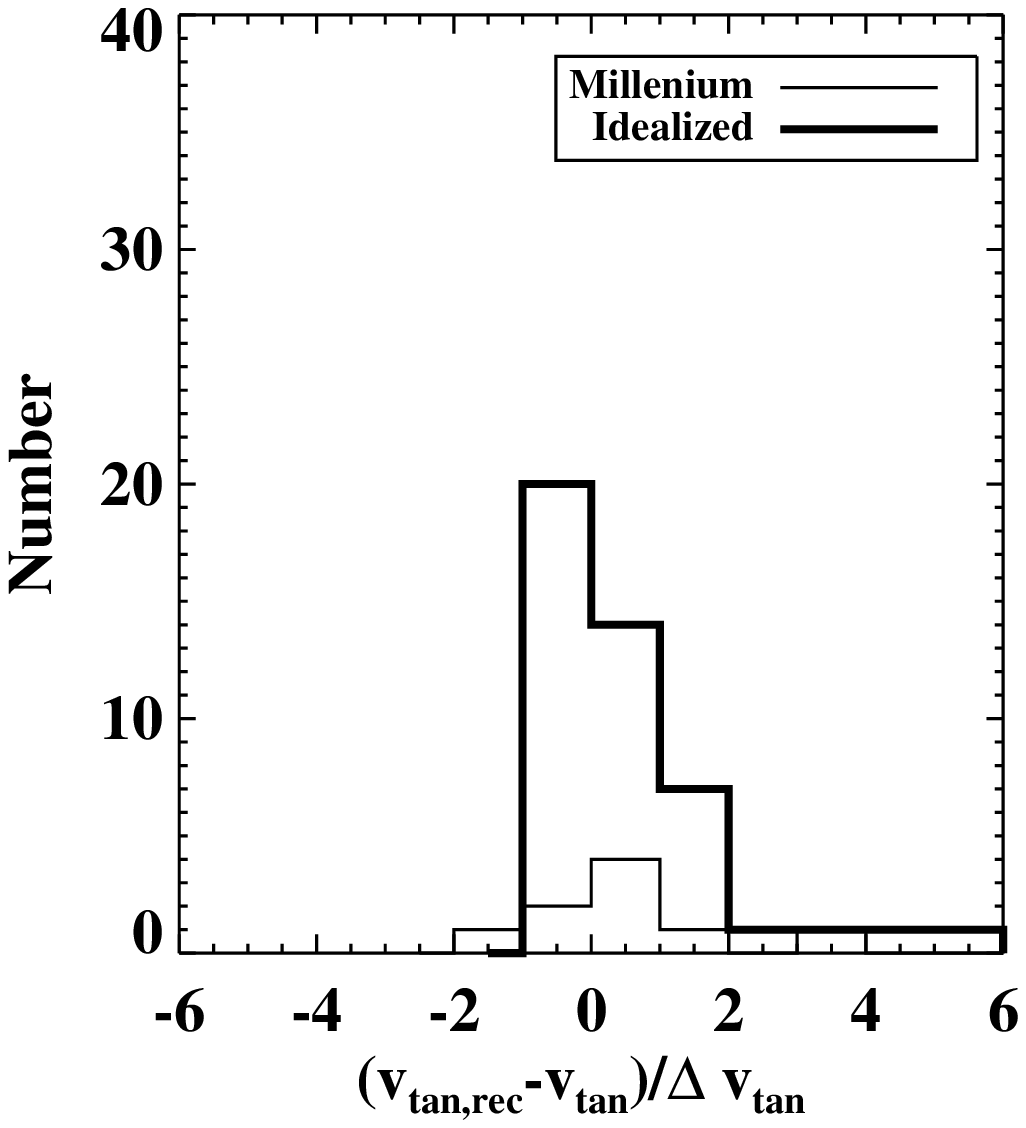}{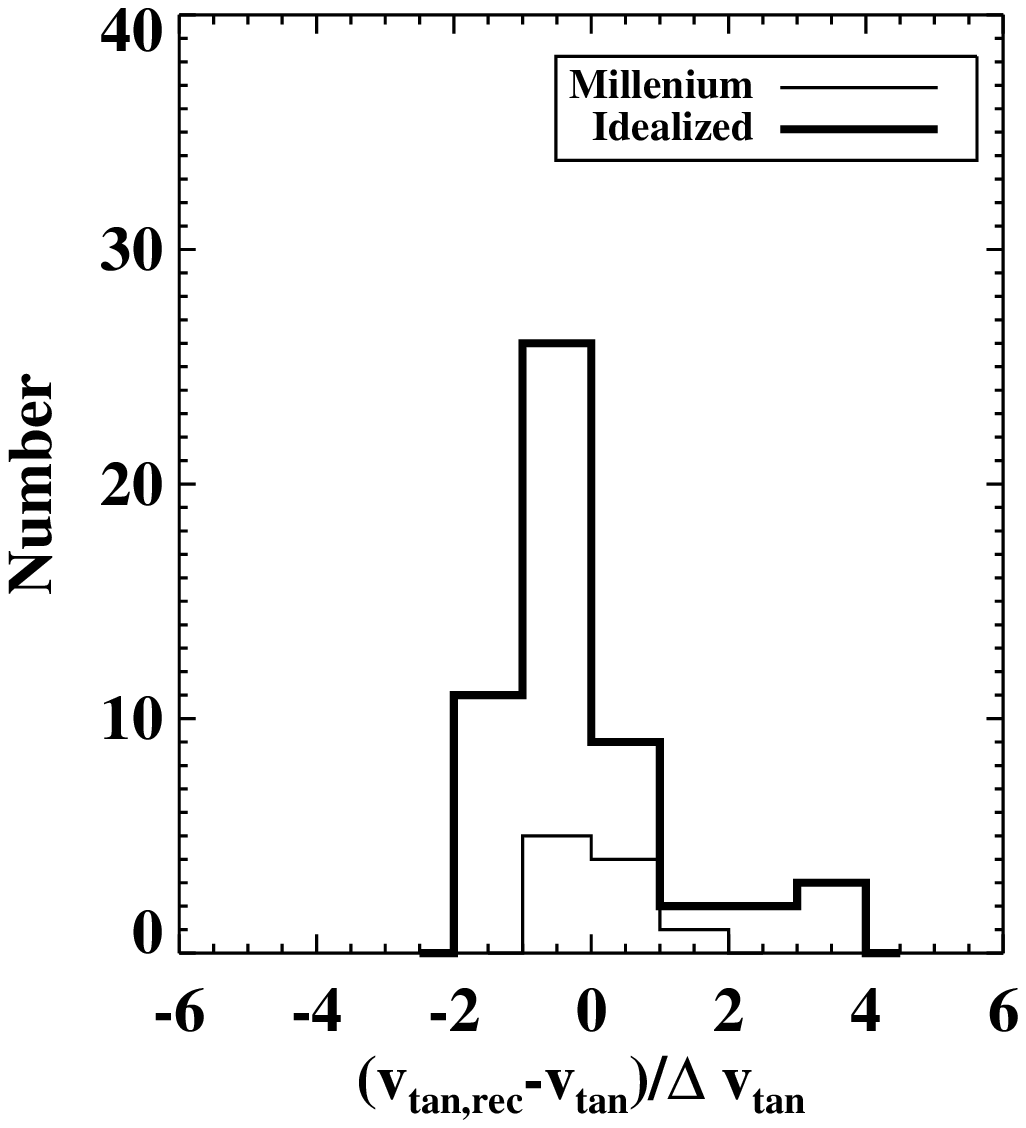}
\caption{Histograms of accuracy of results ($(v_{\rm tan, rec}-v_{\rm tan})/\Delta v_{\rm tan}$) in the north (left panel) and west (right panel)  directions for our 60 idealized  data sets (bold line), and the 60 samples drawn from the Millenium simulated galaxy catalogue (thin line).}
\label{fig:histo}
\end{figure}

The most obvious approach to using PR to find the transverse motion of galaxy clusters is simply to acquire as many line-of-sight velocity measurements of galaxies as possible. 
We tested this idea by randomly generating the projected positions and line-of-sight velocities for sample of $N=500$ galaxies in an idealized, spherically symmetric galaxy cluster, moving with $v_{\rm sys,los}=720$km/s (consistent with the Hubble flow at the distance of the Virgo cluster) and the two components of $v_{\rm sys, tan}$ chosen at random from a gaussian with $\sigma_{\rm pec}=400$km/s. 
The galaxies were uniform in projected density
out to radius  $\theta_{\rm max}=5^\circ$
and had an isotropic gaussian velocity distribution
with dispersion $\sigma=400$km/s.
The unknowns ($v_{\rm sys, rad}, v_{\rm sys,tan}$ and $\phi_{\rm 0}$) in equation (\ref{eq:LOSvel}) were solved for by finding the minimum of $\Sigma_i (v_{i, \rm los} -v_{\rm los}(\theta_i,\phi_i))^2$ 
 in a simple grid-based search of parameter space (where subscript $i$ indicates the value ``observed'' for an individual galaxy). 
By repeating this experiment for many difference sets, we found our uncertainty in the tangential velocity was (as expected) of order:
\begin{equation}
\Delta v_{\rm tan} \sim{1 \over \sin (\theta_{\rm max})} {\sigma \over \sqrt{N}} \sim  {5^\circ \over \theta_{\rm max}} \sqrt{100 \over N} \sigma.
\label{delv}
\end{equation}
The bold solid histogram in Figure \ref{fig:histo} shows the distribution  of difference between our recovered and input tangential velocities ($v_{\rm tan, rec}-v_{\rm tan}$) normalized by the expected error ($\Delta v_{\rm tan}$) for a set of 60 experiments on our idealized data sets. 
Note that the method for bootstrapping estimates of errors from observed data sets adopted by \citet{2008VanDerMarel} was found to agree well with this estimate.

\begin{figure}
\includegraphics[width=0.5 \textwidth]{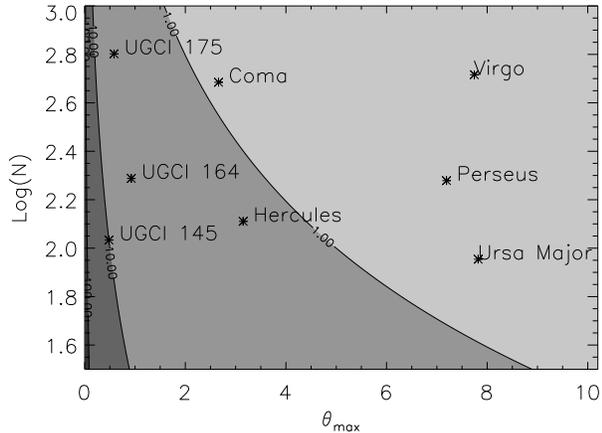}
\caption{Contours of $f=\Delta v_{\rm tan}/\sigma$ as a function of number $N$ and angular extent $\theta_{\rm max}$ of measurements of line-of-sight velocities made.}
\label{fig:nrho}
\end{figure}
Using our results so far we can assess which real galaxy clusters may be interesting targets for a study of PR --- i.e. those for which our
estimated uncertainty ($\Delta v_{\rm tan}$) is less than the expected signal ($\sim \sigma_{\rm pec}$ --- which is comparable to $\sigma$ for galaxy clusters).  
Figure \ref{fig:nrho} shows contours of $f=\Delta v_{\rm tan}/\sigma$ for objects of various $\theta_{\rm max}$ and $N$, with current values for some nearby galaxy clusters overlaid. 
It is clear that the number of measurements for these objects is just becoming interesting for our purpose --- those that have $f \le1$ (i.e. within lightest-gray area of the plot) should have errors due to random motions of order 
$\sigma$ which is similar in amplitude to what we are trying to measure. 

Nevertheless, a  straightforward application  of our grid-based search
to the  $N=379, \theta_{\rm max}=3.4^{\circ}$ Virgo galaxies selected by
\citet{rines08} from the Sloan Digital Sky Survey spectroscopic data
base, and augmented to $N=520, \theta_{\rm max}=7.7^\circ$ with the
\citet{binggeli87} sample, found a tangential velocity of several
thousand km/s with even larger error bars. We interpret this as a null
result, most likely due to systematic motions of groups of galaxies
falling in to Virgo. Indeed, Virgo is classified as an ``irregular''
galaxy cluster for just this reason \citep[e.g.][]{bohringer94}.  

When we applied the search to the Ursa Major cluster using a sample with
$N=90, \theta_{\rm max}=7.8^{\circ}$ galaxies culled from \citet{tully96} and \citet{trentham01} 
we found $v_{\rm sys}=919\pm 39$ km/s, $v_{\rm North}=370 \pm 556$ km/s, $v_{\rm West}=980 \pm 794$ km/s and $\sigma=378 \pm 30$ km/s \citep[where the errorbars are from the bootstrap method outlined in][]{2008VanDerMarel}.
Unfortunately, while the measurements in this case fall within our expectations for clusters' peculiar velocities, this simple interpretation is unlikely to be correct.
Since we live in the epoch of galaxy cluster formation, our problem with Virgo (contamination of the sample by substructure) is likely to be one shared by other galaxy clusters.
In order to assess how significant this concern is, we compared the known space-motions of galaxy clusters in the semi-analytic galaxy catalogue produced for the Millenium Simulation \citep{croton06}, with those we derived by ``observing'' the  same galaxy clusters from random viewpoints, at a distance of 10Mpc.
In all, we analyzed 60 sets of simulated observations of galaxy clusters, with angular sizes of order $\theta_{\rm max}\sim 5^\circ$ and numbers of galaxies in each in the range $N\sim 250-700$.
The dotted histogram in Figure \ref{fig:histo} shows our results --- we find only $\sim$ 10 \% of our experiments give estimates within our expected uncertainty due to random motions. The remaining $\sim 90$\% fall outside the plot. 


While these results are somewhat discouraging, the sizes of the real and simulated catalogues of spectra are still much smaller --- by more than an order of magnitude --- than the actual numbers of galaxies in clusters.
By taking spectra of fainter cluster members
it still may be possible to gather sufficient data  that systematic internal motions (substructure and rotation) can be separated from PR.
Moreover, these samples could be augmented by intra-cluster globular clusters, planetary nebulae and giant stars.

\subsection{.... cluster gas}

The collisional nature of gas suggests that the intra-cluster medium (ICM) could provide a cleaner probe of PR than cluster galaxies --- naively, the ICM might be expected to follow a more symmetric distribution of
velocities around the system mean, and to have lower random local motions (i.e. smaller $v_{\rm local}$ term in equation \ref{eq:LOSvel}).
The measurement of gas velocities from X-ray spectroscopy of the ICM is just becoming feasible: for example \citet{dupke06} report direct detections of velocity differences from X-ray observations of the ICM across the Centaurus galaxy cluster  of $~$2000 km/s. 
Indeed, the planned International X-ray Observatory (IXO) is aiming to have the capability of surveying galaxy cluster gas at high spectral resolution ($\Delta E \sim 2.5$eV for $E \sim 0.3-7$keV on few arcminute scales) in order to study turbulent motions of cluster gas with 100 km/s resolution at $z\sim 2$ \citep{arnaud09}. 
While PR, with an expected signal of order $\sim 100$km/s (see equation \ref{eq:signal}) would be barely detectable in the largest nearby galaxy clusters with this resolution, the IXO design does at least demonstrate the technical feasibility of these measurements.
Unfortunately, the interpretation of any such observations is likely to be difficult: numerical simulations do imply the existence of long-lived turbulent velocities and bulk flows  of order a 300-600 km/s on scales of 100-500kpc in the typical ICM \citep{norman99},
and a comparable degree of rotational support throughout the inner parts of the cluster \citep[as seen in][although these results may in part be due to numerical effects such as over-cooling]{fang09}. Indeed, numerical studies suggest that the rotation of gas following mergers of clusters may be longer lived than for galaxies \citep{roettiger00}.

\subsection{.... the kinetic Sunyaev-Zeldovich effect}

Another way to measure galaxy cluster motions is to look for distortions of the Cosmic Microwave Background (CMB) due to the cluster's intervening gas. 
\citet{sunyaev72} demonstrated how inverse compton scattering of CMB photons from energetic electrons in the ICM could be 
detectable as a (frequency dependent) temperature change. This effect --- commonly referred to as the {\it thermal} or {\it static} Sunyaev-Zeldovich effect --- has since become a standard tool for analyzing ICM properties \citep[e.g.][and references therein]{muchovej07}
\citet{sunyaev80} also noted that 
any motion of the ICM relative to the CMB rest frame would impart an additional (frequency independent) doppler distortion to the temperature of the CMB --- dubbed the {\it kinetic} Sunyaev Zeldovich effect (kSZ) --- of order
\begin{equation}
{\Delta T \over T}=- {v_{\rm los} \over c} \tau
\end{equation}
where $\tau$ is the optical depth with respect to Thomson scattering.
The current generation of CMB experiments should be  able to use the kSZ effect to successfully measure the peculiar motions of galaxy clusters along the line-of-sight \citep[see for e.g.][]{cunnama09}, which are expected in general to produce maximum distortions $\Delta T \sim 20 \mu$K \citep{molnar00}.
PR induced by the galaxy cluster's transverse motion would be apparent as a dipole signature in the kSZ temperature distortion. The amplitude of this distortion would be smaller than the typical line-of-sight velocity signature by two factors: the first of order $\sin \theta_{\rm max}$ (or 5-10\%  for nearby galaxy clusters) due to the projection effect; and the second due to the fall-off in gas-density (and corresponding decrease in $\tau$) away from the center of the galaxy cluster.
Hence, we might expect PR to induce 
a distortion of order
 0.1-1 $\mu$K across nearby galaxy clusters --- clearly a challenge for near-future experiments.
As before, the signal of PR in the kSZ effect will also be competing with signatures of turbulent motions \citep[expected to be of order 10$\mu$K, see][]{sunyaev03} and intrinsic rotation \citep[expected to be of order 2$\mu$K due to tidally induced rotation alone][]{cooray02}.


\section{Summary, discussion and conclusion.}

We have outlined three possible ways to measure the PR of galaxy clusters and estimate their transverse motions: from line-of-sight velocity measurements of galaxy cluster galaxies, the motion of cluster gas and mapping the kSZ distortions in the CMB.
All three of these measurements are unfeasible with the current data but could become feasible with near future instruments or larger data sets.
The amplitude of PR is most significant for galaxy clusters with larger angular extent.
Hence, the most promising approach with current capabilities is to survey as large a sample as possible of spectra of galaxies in nearby galaxy clusters.

With all approaches, the signal of PR could be confused by random (or turbulent) motions and intrinsic rotation of their system of targets.
We are optimistic that it will be possible to disentangle these effects with sufficiently large
data sets since
PR imposes a unique signature of solid-body rotation on top of these sources of confusion.
Indeed, \citet{2008Kaplinghat} found that, when they included a small intrinsic rotation in
their models of dwarf spheroidal galaxies, 
the error bars on the transverse velocity estimates derived by ``observing'' line-of-sight velocities of stars in their models increased
by only a factor of two.
While galaxy clusters are unlike dwarfs spheroidal galaxies in that they are not expected to
be as relaxed or spherically symmetric, the \citet{2008Kaplinghat} study provides
a first step towards a comprehensive modeling effort.

Note that several other approaches to detecting transverse motions of
galaxy clusters have also been proposed. For example using polarization maps of the CMB \citep{sunyaev80}, gravitational lensing of the CMB \citep{birkinshaw83}, and weak and strong lensing of background galaxies \citep{molnar03}. The strength of the signatures in these methods are not dependent on the angular extent of the galaxy cluster and hence have the advantage over PR of being applicable to distant as well as nearby clusters.

With multiple possible directions for detection, we conclude that measurements of the full space motions of galaxy clusters are on the horizon.

\begin{acknowledgements}

The authors wish to thank Greg Bryan for keeping us correct, honest and realistic, Zoltan Haiman for directing us towards the CMB and Andy Gould and Avi Loeb for general encouragement.

The Millennium Run simulation used in this paper was carried out by the Virgo Supercomputing Consortium at the Computing Centre of the Max-Planck Society in Garching. The semi-analytic galaxy catalogue is publicly available at \\
http://www.mpa-garching.mpg.de/galform/agnpaper

This work was supported by NSF grant AST-07-34864.

\end{acknowledgements}


\end{document}